# Oxidation of CO on surface hematite in high $CO_2$ atmospheres

John Lee Grenfell[1]*, Joachim W. Stock[2], A. Beate C. Patzer[1],

Stefanie Gebauer[1], and Heike Rauer[1,2]

(1) Zentrum für Astronomie und Astrophysik

Technische Universität Berlin (TUB)

Hardenbergstr. 36

10623 Berlin

Germany

(2) Institut für Planetenforschung

Deutsches Zentrum für Luft- und Raumfahrt (DLR)

Rutherford Str. 2

12489 Berlin

Germany

*Corresponding Author: John Lee Grenfell

Email: lee.grenfell@dlr.de

Telephone: +49 30 314 25463

27

28

29 **Abstract**: *We propose a mechanism for the oxidation of gaseous CO into $CO_2$*

30 *occurring on the surface mineral hematite ($Fe_2O_3(s)$) in hot, $CO_2$-rich planetary*

31 *atmospheres, such as Venus. This mechanism is likely to constitute an important*

32 *source of tropospheric $CO_2$ on Venus and could at least partly address the $CO_2$*

33 *stability problem in Venus' stratosphere, since our results suggest that atmospheric*

34 *$CO_2$ is produced from CO oxidation via surface hematite at a rate of 0.4*

35 *Petagrammes (Pg) $CO_2$ per (Earth) year on Venus which is about 45% of the mass*

36 *loss of $CO_2$ via  photolysis in the Venusian stratosphere. We also investigated CO*

37 *oxidation via the hematite mechanism for a range of planetary scenarios and found*

38 *that modern Earth and Mars are probably too cold for the mechanism to be important*

39 *because the rate-limiting step, involving CO(g) reacting onto the hematite surface,*

40 *proceeds much slower at lower temperatures. The mechanism may feature on*

41 *extrasolar planets such as Gliese 581c or CoRoT-7b assuming they can maintain*

42 *solid surface hematite which e.g. starts to melt above about 1200K. The mechanism*

43 *may also be important for hot Hadean-type environments and for the emerging class*

44 *of hot Super-Earths with planetary surface temperatures between about 600-900K.*

45

46

47

48

50

## 51 **1. Introduction**

52

53     Venus and Mars have maintained rich $CO_2$ atmospheres over geological periods
54 despite loss of this molecule via photolysis. How $CO_2$ is re-generated from its
55 photolysis products is usually termed the "$CO_2$ stability problem." On Mars, catalytic
56 photochemical cycles can mostly account for this re-regeneration (McElroy and
57 Donahue, 1972; Parkinson and Hunten, 1972; Yung and DeMore 1999). On Venus,
58 however, the situation is less clear. The effectiveness of the proposed Venus reaction
59 cycles is not well-determined, e.g. because rate data of important chemical reactions
60 are lacking for a $CO_2$ bathgas and are challenging to obtain under Venus conditions.
61 Further, key intermediate species such as the chloroformyl radical, essential for the
62 proposed chlorine catalytic cycles (Yung and DeMore, 1999; Pernice et al. 2004) have
63 still to be detected in-situ in Venus' atmosphere. Improved understanding of the
64 photochemistry of $CO_2$-rich atmospheres will also aid in the future assessment of
65 exoplanetary spectral signatures of potential biomarkers i.e. life-indicating species.

66     In this paper we discuss an alternative mechanism to account for the stability of
67 hot, high-$CO_2$ atmospheres, which proceeds via oxidation of CO occurring on
68 hematite on the planet's surface. The basic process is well-known in the chemical
69 industry (e.g. Strassburger, 1969). We estimate the rate of $CO_2$ production via this
70 mechanism for a range of different planetary scenarios such as Venus, Earth, Early
71 Earth, Mars, and hot exoplanet conditions.

72

73

74

75

## 76 **2. Method**

77

### 78 **2.1 Heterogeneous catalysis on hematite**

79

80      $CO(g)$ and $O_2(g)$ adsorb quickly onto many iron-containing surfaces (Atkins,

81 1986) via chemisorption e.g. between the oxygen 2p orbitals of CO and the 3d orbitals

82 of iron in the surface of hematite (e.g. Becker et al., 1996; Föhlisch et al., 2000;

83 Reddy et al., 2004). The adsorbed species' bonds are weakened and can react to form

84 $CO_2$. For CO oxidation on hematite, there are some indications that both the Eley-

85 Rideal mechanism (ER), i.e. where an adsorbed species reacts with a gas-phase

86 species (Atkins, 1986) and the Langmuir-Hinshelwood mechanism (LH) i.e. where

87 both reacting species must first adsorb, can occur. The ER mechanism for CO on

88 hematite was discussed e.g. by Halim et al., (2007), and Wagloehner et al. (2008),

89 whereas the LH mechanism was discussed by Bergermayer and Schweiger, (2004);

90 Kandalam et al., (2007) discussed both mechanisms. There are numerous theoretical

91 studies which are investigating this issue (e.g. Panczyk, 2006; Bulgakov and Sadykov,

92 1996; Reddy et al., 2004).

93      Figures 1a-c illustrates the various steps in the mechanism adopted in our

94 work for the oxidation of CO on hematite. Hematite particles consist of a bulk (dark

95 shading) and surface (light shading) region. Active adsorption sites ("holes") exist on

96 the surface (Randall et al., 1997; Wagloehner et al., 2008: Bergmayer et al., 2004),

97 some of which may be occupied by trapped oxygen atoms, which can originate either

98 via diffusion from the crystal bulk, through the lattice to the surface region (e.g.

99 Kandalam et al., 2007; Randall et al., 1997) or from the atmosphere directly via

100 dissociative adsorption of e.g. $O_2(g)$ or/and $CO_2(g)$. Migration across the surface is not

101 considered to be significant for CO oxidation on hematite because adsorbed species

102 are confined to the active sites (Randall et al., 1997).

103       The first step (Figure 1a) involves $O_2$ adsorption. This process is suggested to

104 involve the radical anion reactive intermediates $O^-$ or $O_2^-$ with different adsorbed

105 states depending on how the $O_2$ approaches the surface e.g. perpendicularly, sideways

106 or obliquely (Kandalam et al., 2007; Bergermayer and Schweiger, 2004). Also

107 important is the number of surface iron atoms, with which the adsorbed $O_2$ can

108 interact (Bulgakov and Sadykov, 1996). The next step (Figure 1b) involves gaseous

109 CO removing the adsorbed oxygen atom. The final step (Figure 1c) involves

110 desorption of the $CO_2$ product leaving behind an active site occupied with an adsorbed

111 oxygen atom.

112       Details of the individual mechanism steps are, however, not certain. The

113 theoretical study of Kandalam et al. (2007), for example, suggested that a CO

114 molecule first adsorbs onto hematite, weakening an Fe-O bond near the crystal

115 surface. Then, a second CO molecule adsorbs and forms $CO_2$ by breaking the

116 weakened Fe-O bond. Further reaction rate data are needed, however, to assess this

117 particular mechanism. In our study we focus on the ER mechanism illustrated in

118 Figure 1a-c, for which detailed kinetic data are available (Wagloehner et al., 2008).

119

120 **2.2 Experimental data for oxidation of CO on hematite**

121

122       Obtaining experimental rate data is challenging mainly because the

123 morphology and size-distribution of the hematite crystal, hence the CO oxidation rate,

124 can vary from experiment to experiment (Lin et al., 2005; Kwon et al., 2007). Such

125 chemical rates are determined by passing carrier gases containing typically a few

126 percent of CO and/or $O_2$ over hematite powder then measuring the conversion rate of 127 CO into $CO_2$ based on mass spectroscopy, for a range of temperatures and pressures. 128 Typical data suggests between (10-100)% conversion of the CO by mass into $CO_2$ per 129 second at normal laboratory flow rates of about 1 litre per minute, typically with 130 (0.1-1.0)g powdered hematite at (400-700)K (e.g. Tripathi et al., 1999; Khedr et al., 131 2006; Li et al., 2003; Khoudiakov et al., 2004; Randall et al. 1997). The overall rate 132 constant ($k_i$) features the well-known Arrhenius dependence upon temperature, such 133 that $k_i= A_i exp(-E_{a,i}/(RT))$ where $A_i$ denotes the pre-exponential constant; $E_{a,i}$ is the 134 activation energy; R is the gas constant and T denotes the temperature.

135        In our study we use the detailed kinetic data based on the investigation of 136 Wagloehner et al. (2008), which consists of the following rate expressions (see also 137 Figures 1a-c):

138                    $R_1 = A_1 exp(-E_{a,1}/RT) \, [O_2]\theta_*^2$

139                    $R_2 = A_2 exp(-E_{a,2}+\alpha 2\theta_O)/(RT)\theta_O^2$

140                    $R_3 = A_3 exp(-E_{a,3}/RT)[CO]\theta_O$

141                    $R_4 = A_4 exp(-E_{a,4}-\alpha 4\theta_O)/(RT)\theta_{CO_2}$

142                    $R_5 = A_5 exp(-E_{a,5}/RT)[CO_2]\theta_*$

143                    $R_6 = A_6 exp(-E_{a,6}+\alpha 6\theta_O)/(RT)\theta_{CO_2}$

144

145 Where $R_i$ denotes the reaction rate of reaction, i (mol m$^{-3}$ s$^{-1}$), [X] is the gas phase 146 concentration of species X, (mol m$^{-3}$); $\theta_X$, the  fractional coverage of hematite surface 147 by species X; $\alpha_*$, a constant and $\theta_*$, the fractional coverage of unoccupied, active sites 148 on the hematite surface. The quadratic theta terms in $R_1$ and $R_2$ arise because we 149 follow the arguments of Wagloehner et al. (2008), who assumed the adsorbing $O_2$

150 binds to *two* active sites. All rates are multiplied by the surface area of hematite per

151 unit volume.

152       In addition we parameterised the diffusion rates of O-atoms from the internal

153 hematite bulk to the surface ($R_{out}$) and from the surface back into the bulk ($R_{in}$) based

154 on rate data from Randall et al. (1997), using:

155

156                     $$R_{in} = R_{out} = A_{diffusion} e(-E_{diffusion}/(RT))$$

157

## 158 2.3 Computational details

159

160       The chemical rate expressions discussed in section 2.2 were integrated using

161 the FACSIMILE program, in which non-linear, time-dependent chemical reactions are

162 treated as source terms and solved as a system of differential equations using Gear's

163 method (Gear et al. 1985). The differential equation system contains eight rate

164 equations, i.e. $R_1$-$R_6$, $R_{in}$ and $R_{out}$, as discussed in section 2.2 for the variables: $O_2(g)$,

165 $CO(g)$, $CO_2(g)$, $\theta^*$, $\theta_O$, $\theta_{CO}$, and $\theta_{CO2}$. The solution of this system with appropriate

166 initial values yields their temporal development approaching the equilibrium situation.

167       Wagloehner et al. (2008) assumed *constant* coverages of $\theta_O = 0.6$, $\theta_{CO2} = 0.4$,

168 and $\theta_* = 4 \times 10^{-5}$ in their kinetic modelling experiment, which was sufficient to

169 reproduce their laboratory-observed $CO_2$ formation rate. In our study, however, since

170 we wish to apply our model flexibly to a range of planetary environments, we

171 modified the Wagloehner et al. concept so that surface coverage is not considered as

172 given, fixed values, but instead is calculated *variably* in our model. With our variable

173 scheme, our modelled results were consistent with the observed CO oxidation rates of

174 Wagloehner et al. (T=265$^o$C, P=1bar, 0.5g hematite sample with surface area of

1757.5m$^2$), taking a hematite active site surface density of $3.7 \times 10^{18}$ m$^{-2}$ based on
176 Iwamoto et al. (1978). Results from our interactive model, showing CO and $CO_2$
177 concentrations for the above-mentioned observed laboratory conditions of the
178 Wagloehner et al. study, are shown in Figure 2a, whereby CO is completely oxidised
179 into $CO_2$ on hematite within timescales of a few hours. Figure 2b shows surface
180 coverage variables calculated in our model for the laboratory conditions. Results
181 imply that our interactive $\theta_O$ and $\theta_{CO_2}$ values differ from the prescribed values used by
182 Wagloehner et al. by about 50% mainly because, we are modelling a mass-isolated
183 box, whereas the Wagloehner et al. experiment featured a constant supply of reagant
184 gas being pumped over the hematite surface.

185        To be applied to planetary atmospheres, we consider a local volume element
186 located at the surface of the planet with – where possible – observed values for the
187 prevailing chemical and physical conditions i.e. the hematite surface coverage, the
188 temperature, the pressure, the atmospheric CO(g) and $O_2$(g) content. We assume zero
189 atmospheric $CO_2$ at the initial time of the integration. Starting with zero particle
190 coverage, we solve the differential equations by forward time integration until the
191 $CO_2$ concentration reaches a steady-state value (c.f. Figures 4a, 4b). To calculate the
192 global $CO_2$ production, the resulting $CO_2$ production rate ($R_3$, mol m$^{-3}$ s$^{-1}$) is multiplied
193 by the surface area of the planet assuming a smooth sphere.

194

195 **2.4. Boundary Conditions for the Planetary Scenarios**

196

197        The various conditions of the seven planetary scenarios conditions are
198 summarised in Table 1. For modern Venus, several studies have suggested the
199 presence of hematite on the surface (e.g. Fegley et al., 1995; Klingelhöfer and Fegley,

200 2000; Roatsch et al., 2008). However, some works have questioned whether magnetite

201 may be more stable at the surface (Wood, 1996), with hematite favoured near

202 mountain tops (Fegley et al., 1997). Other works suggested that hematite could be

203 present in a nano-phase form on Venus (Straub et al., 1991; Straub and Burns, 1990).

204 In the following we assume a 1% surface coverage of hematite, which is the lower

205 limit suggested by Klingelhöfer and Fegley (2000) and an amount of observed $O_2(g)$

206 equal to the upper limit of $(3 \times 10^{-7})$ volume mixing ratio (vmr) in Venus' atmosphere

207 (Trauger and Lunine, 1983).

208

209 **3. Results**

210

211        Table 2 shows the global rate of $CO_2$ production in $10^{15}$g yr$^{-1}$ (i.e.

212 petagrammes (Pg) $CO_2$ per Earth year) for the seven planetary scenarios. The results

213 suggest that on modern Venus the proposed hematite mechanism produces about

214 0.4Pg $CO_2$/yr at its surface with the assumed 1% hematite coverage.

215        Figure 3 presents an overview of the atmospheric $CO_2$ budget on Venus and

216 implies that the hematite mechanism could be of central importance in the

217 troposphere. Escape rates in Figure 3 are based on Lammer et al. (2008). Boxes

218 marked HOx, SOx and ClOx in the stratosphere of Figure 3 represent the contribution

219 from photochemical catalytic cycles involving hydrogen, sulphur-, and chlorine

220 oxides respectively. Values are calculated from the model study of Yung and DeMore

221 (1999) (chapter 8 and references therein) and have been averaged from 60km up to

222 the stratopause. Figure 3 suggests that potentially fast $CO_2$ production from  SOx and

223 ClOx *may* offset loss of $CO_2$ via photolysis hence address the stability problem.

224 However, there are some important caveats. Firstly, SOx values are uncertain due to

225poorly-defined fluxes of sulphur-containing compounds  from the troposphere to the

226stratosphere. Secondly,  ClOx values are uncertain since reactive chlorine

227intermediates required for such cycles have not been observed in-situ. Mills and Allen

228(2007) discuss these issues in the context of the ongoing stability problem on Venus.

229At the surface, the volcanism contribution in Figure 3 is based on Fegley and Prinn

230(1989) who suggested that $CO_2$ emissions from volcanoes on Venus are weaker than

231on Earth, which emits <0.5 Pg/yr $CO_2$ (Gerlach 1991).

232        To address whether our mechanism at the *surface* is able to address the $CO_2$

233stability problem in the *stratosphere*, Figure 3 shows the global mean rates of Eddy

234diffusion, as calculated via the diffusion equation:

235

236                        Flux = -K(dn/dz + n/H + (1/T)(dT/dz) n )

237

238        In the above, we consider a 1km thick layer at the tropopause located at a

239height of 60km. K represents the Eddy diffusion coefficient, taken to be $(0.8-10) \times 10^4$

240$cm^2 s^{-1}$ (Yung and DeMore, 1982), which is the main source of uncertainty in the

241calculation. Remaining variables were calculated from the Venus International

242Reference Atmosphere (VIRA) database (Seiff, 1983); n is the number density of $CO_2$

243taken to be $4.71 \times 10^{18}$ molecules $cm^{-3}$ at 60km, H is the scale height with the value

2446km at 60km; T is the temperature taken to be 234K at 60km. Results suggest that

245Eddy diffusion is fast and does not limit the supply of $CO_2$ from the troposphere.

246        Figures 4a-b are similar to Figures 2a-b but for modern Venus. Figure 4a

247implies that CO is reduced, being converted into $CO_2(g)$, which rises as shown in

248Figure 4a, and also into $CO_2$(ads) which rises as shown in Figure 4b. For Venus

249conditions, the fractional coverage of vacant active sites ("holes") on the hematite

250surface increases to about 0.5 (Figure 4b) compared with ~$10^{-5}$ for the laboratory

251conditions (Figure 3b). This was due to the higher Venus temperatures, which enabled

252adsorbed species to escape more easily from the hematite surface.

253        We performed a sensitivity study, varying the boundary conditions in our

254model. Lowering atmospheric oxygen in the Venus run by a factor of 10,000 had only

255a small impact on the results, since the supply of adsorbed oxygen atoms was still

256saturated via diffusion from the bulk crystal. Changing the assumed hematite surface

257coverage for Venus impacted the $CO_2$ production rate linearly.

258        Table 2 shows surface coverage data and resulting global $CO_2$ production rates

259(Pg/yr) for the seven planetary scenarios. These results suggest that on modern Earth,

260modern Mars, and the Early Earth scenarios, the hematite mechanism is not important

261due to a slowing in the temperature-dependent chemical rates. For the hypothetical

262Super-Earth scenario, however, the hematite mechanism could be quite important.

263Finally, for the recently discovered hot exoplanets Gliese 581c and CoRoT-7b, the

264hematite mechanism could be important, especially at higher surface pressures, as

265suggested by the high $CO_2$ production in Table 2 for the Gliese 581c scenario (with 93

266bar surface pressure) with the much lower production for the CoRoT-7b scenario

267(with 0.1 bar surface pressure), assuming that the surface is not too hot to melt

268hematite or/and convert all hematite into magnetite, which is more stable at warm

269temperatures.

270In summary, the catalytic formation of CO(g) into $CO_2$(g) on hematite surfaces can
271play an important role in some planetary atmospheres. However, the efficiency of the
272process depends critically on the surface temperature and the amount of surface
273hematite.
274
275
276**4. Conclusions**

277

278        Oxidation of CO on surface hematite may address at least partly the long 279standing $CO_2$ stability problem in Venus' atmosphere- our results imply that $CO_2$ is 280generated  from CO oxidation via the hematite mechanism at about 45% (see Figure 2813) of its rate of mass loss via  photolysis in the Venusian stratosphere.

282        As Early Venus warmed, the hematite mechanism increased, contributing a 283positive feedback to Venus' climate, which could have played an important role in the 284well-known runaway climate scenario.

285        The hematite mechanism may also play an important role in stabilising the 286$CO_2$ atmospheres of hot Super-Earths with surface temperatures in the range of 287(600-900)K. For cooler environments, such as on Earth, Early Earth, or Mars, 288strongly temperature-dependent rates virtually switch off the hematite mechanism, 289whereas in very hot environments the mechanism may also play an important role.

290        $O_2$ and $N_2O$ are central biomarkers whose atmospheric responses have been 291recently studied in a wide range of Earth-like scenarios (e.g. Segura et al., 2003; 292Grenfell et al., 2007). Both species can adsorb onto hematite and other transition 293metal surfaces, undergoing significant chemical change at temperatures typically in 294the range of hot Super-Earths and maybe at even somewhat cooler conditions. For 295example, $O_2$ undergoes dissociative adsorption on hematite, $N_2O$ adsorbs then 296decomposes to release $O_2$. Such effects are currently not considered in models of 297Earth-like atmospheres.

298

299**Acknowledgement**

300We are grateful to Steffen Wagloehner and to Philip von Paris for useful discussion. 301This research has been supported by the Helmholtz Gemeinschaft through the 302research alliance "Planetary Evolution and Life".

303

474
475 Figure Captions
476
477
478 Figure 1a:  *Hematite mechanism step 1*: adsorption and desorption of $O_2$(g) on the
479 hematite surface.
480
481 Figure 1b:  Hematite *mechanism step 2*: reaction and reverse reaction of CO(g) with
482 $O_2$(ads) on the hematite surface.
483
484 Figure 1c: *Hematite mechanism step 3*: desorption and adsorption of $CO_2$(g) on
485 hematite surface.
486
487 Figure 2a: Model concentrations to reproduce Earth laboratory conditions (T=265°C,
488 P=1bar, 0.5g hematite, initially 50% $O_2$(g) $7x10^{-3}$ vmr, CO (g), fill-gas $N_2$).
489
490 Figure 2b: Fractional coverage of hematite surface by O(ads), $CO_2$(ads), and active
491 sites ("holes") under laboratory conditions (see section 3).
492
493 Figure 3: Planetary budget of atmospheric $CO_2$(g) on modern Venus in petagramme
494 (Pg) $CO_2$/Earth year. $HO_x$(g), $SO_x$(g) and $NO_x$(g) denote catalytic cycles arising from
495 hydrogen, sulfur, and nitrogen oxides respectively.  *Assuming $3x10^{-7}$ vmr $O_2$(g) and
496 1% hematite surface coverage.
497
498 Figure 4a: Model concentrations for modern Venus scenario (see Table 1) with
499 $O_2$(g)=$3x10^{-7}$ volume mixing ratio (vmr).
500
501 Figure 4b: Fractional coverage of hematite surface by O(ads), $CO_2$(ads) and active
502 sites ("holes") for modern Venus scenario as for Figure 4a.
503
504
505
506
507
508
509
510
511
512
513
514
515
516
517
518
519
520 Figure 1a

## O$_2$(g) adsorbs to active sites

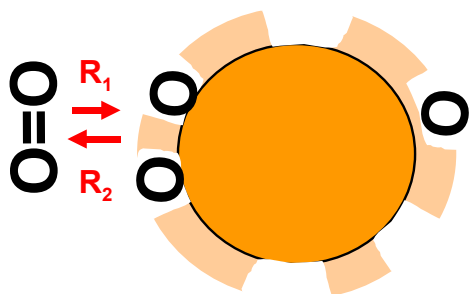

521
522 Figure 1b
523

## CO(g) reacts with O(ads)

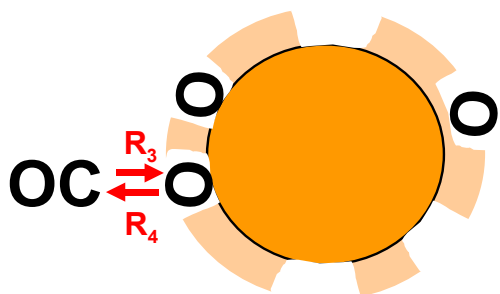

524
525 Figure 1c
526

## CO$_2$ desorbs

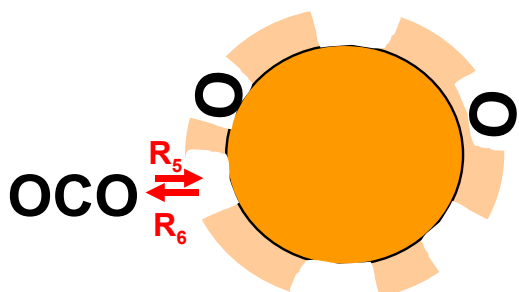

527
528
529
530 Figure 2a
531

532

ontrol: Concentrations

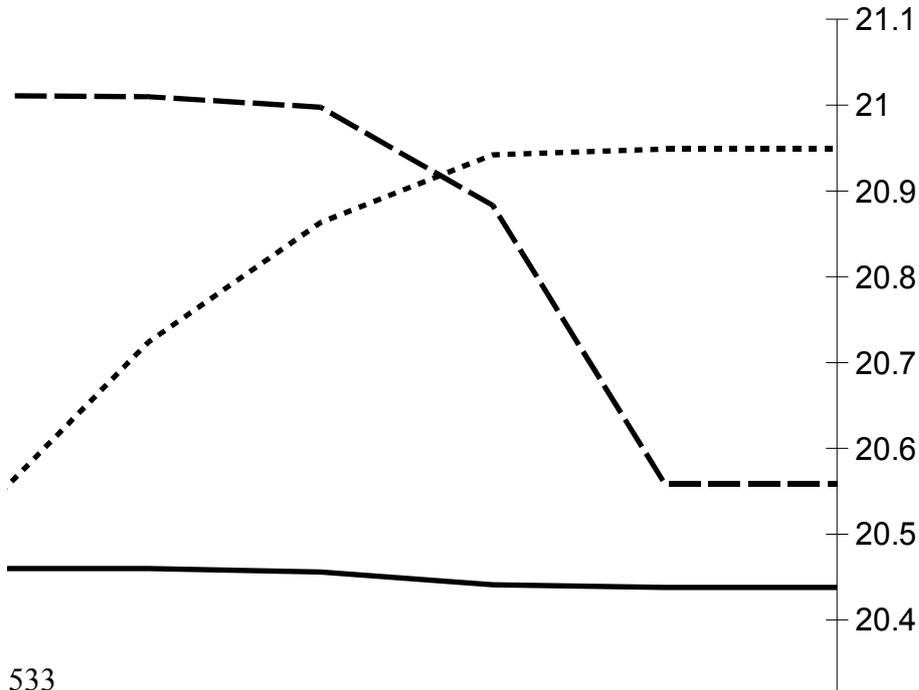

533
534Figure 2b
535

ontrol: Hematite Surface

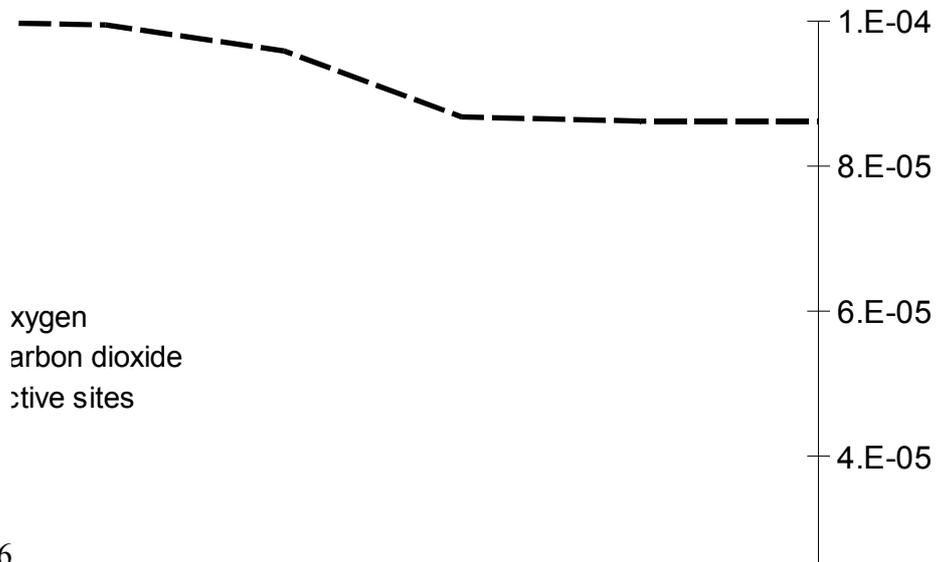

xygen
arbon dioxide
ctive sites

536
537
538
539
540
541
542
543

544
545
546 Figure 3
547
548

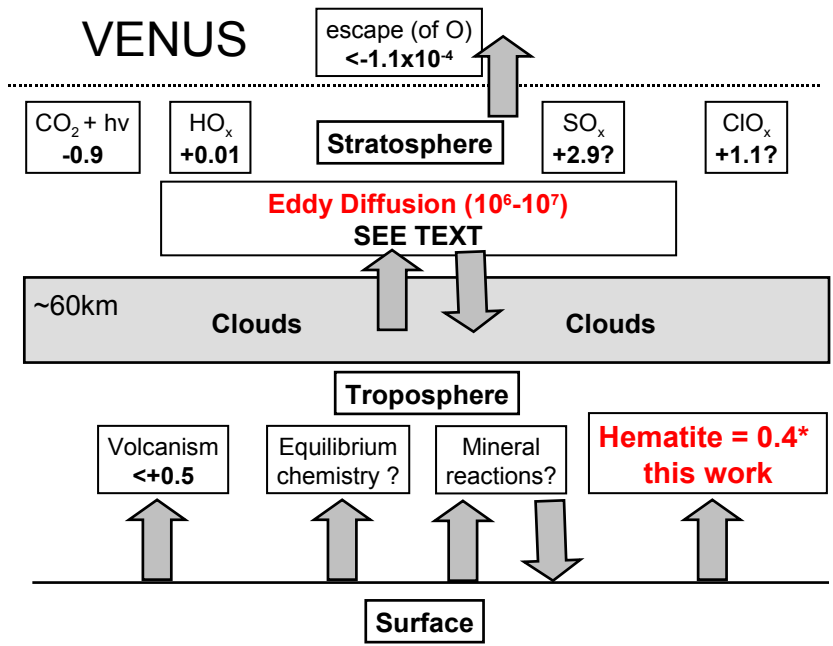

549
550
551
552
553
554
555
556
557
558
559
560
561
562
563
564
565
566
567
568
569
570
571
572
573
574 Figure 4a
575

## Venus: Concentrations

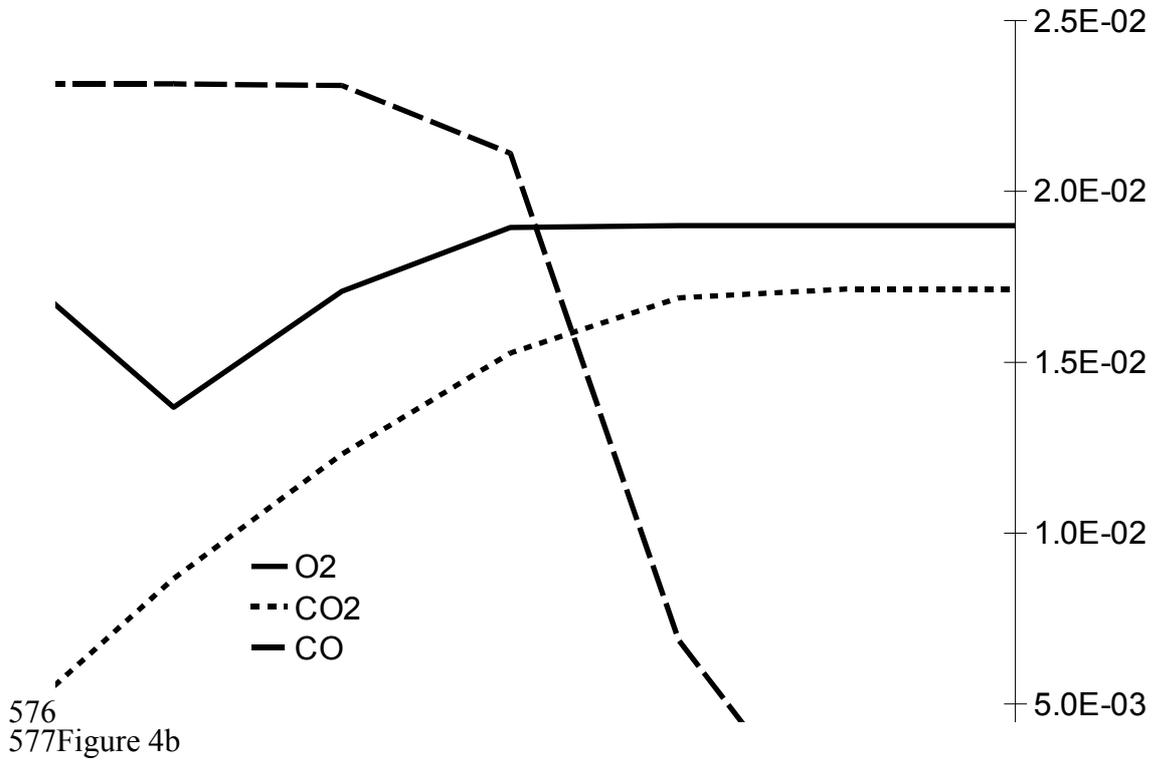

576
577 Figure 4b

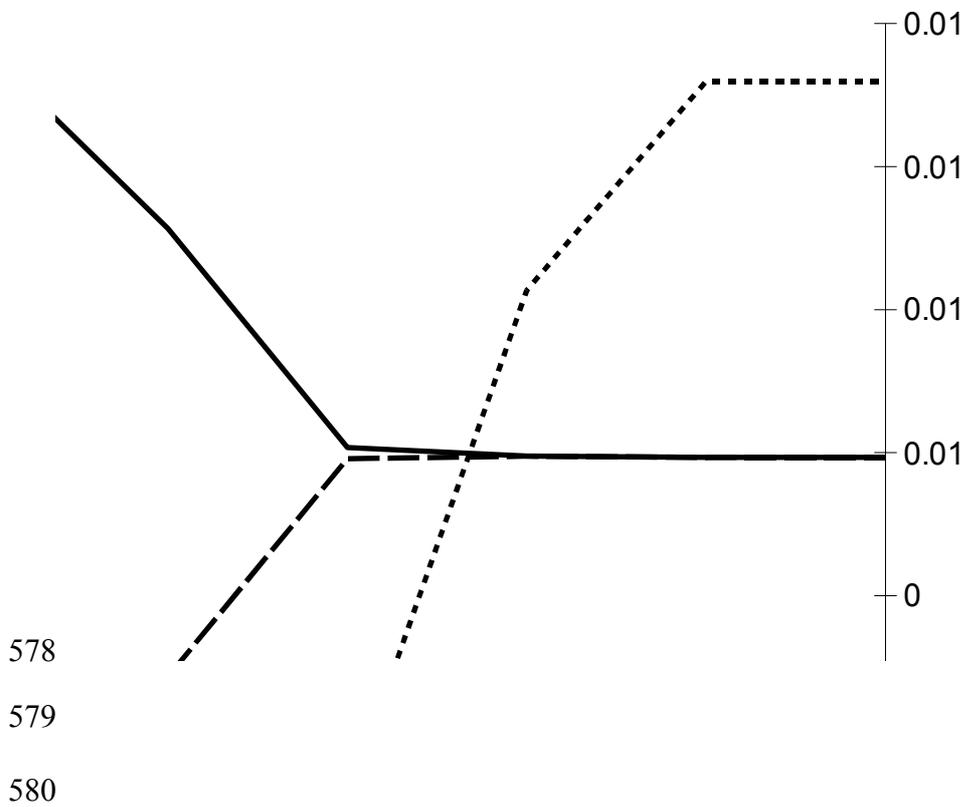

578

579

580

581 Table 1: Input data for planetary scenarios considered.
582
583

| Scenario | Hematite (% coverage at surface) | $T_0$ Surface (K) | $P_0$ Surface (bar) | $[CO]_0$ surface volume mixing ratio | $[O_2]_0$ surface volume mixing ratio | Mean Planetary Radius (R) (km) |
|---|---|---|---|---|---|---|
| Modern Earth | 2.6 (Clarke, 1924[#]) | 288 | 1.0 | $1.25 \times 10^{-7}$ (Yung and DeMore 1999) | 0.21 | $R_{Earth}$ =6371.009 |
| Modern Venus | 1.0- (Klingelhöfer and Fegley, 2000) | 735 (Yung and DeMore 1999) | 93 (Seiff et al. 1983) | $1.5 \times 10^{-5}$ Krasnopolsky (2007) | $<3 \times 10^{-7}$ Trauger and Lunine (1983) | 6051.8 |
| Modern Mars | 3.0 (Encrenaz et al. (2004) | 220 (Yung and DeMore 1999) | $5.6 \times 10^{-3}$ (Yung and DeMore 1999) | $17 \times 10^{-4}$ (Owen et al. (1977) | $1.3 \times 10^{-3}$ (Owen et al. (1977) | 3389.5 |
| Hot Archean | 2.6 | 343 (Robert and Chaussidon 2006) | 1.0 | $8.0 \times 10^{-5}$ (Kasting and Catling, 2003) | $\sim 10^{-12}$ times modern (Kharecha et al. 2005) | $R_{Earth}$ |
| Hot Super-Earth | 1.0## | 800## | 1.0## | As for Venus | As for Venus | $2.0 R_{Earth}$ |
| Gliese 581c | 1.0## | 1000 (Selsis et al. 2007) | 93## | As for Venus | As for Venus | $1.5 R_{Earth}$ Udry et al. (2007)### |
| CoRoT-7b | 1.0## | 1000 (Léger et al. 2009) | 0.1## | As for Venus | As for Venus | $1.68 R_{Earth}$ Léger et al. (2009) |

584
585 #=mass hematite in Earth's crust; ## no observations exist – sensitivity value only;
586 ### assuming an earth composition.
587
588
589
590
591
592
593
594
595
596
597
598
599
600
601
602
603

604 Table 2: Fractional coverage of hematite with O-atoms, $CO_2$-atoms, and vacant active
605 sites, resulting $CO_2$ production for the hematite mechanism at the planet surface for
606 the seven scenarios.
607
608

| Scenario | $\theta_O$ | $\theta_{CO_2}$ | $\theta_{active sites}$ | $CO_2$ (Pg/yr) |
|---|---|---|---|---|
| Laboratory | 0.88 | 0.12 | $1.30 \times 10^{-5}$ | not applicable |
| Venus | 0.49 | $1.11 \times 10^{-2}$ | 0.49 | 0.38 |
| Earth | 1.00 | $3.56 \times 10^{-12}$ | $1.62 \times 10^{-10}$ | $2.49 \times 10^{-13}$ |
| Mars | $8.07 \times 10^{-4}$ | $6.1 \times 10^{-20}$ | 1.00 | $9.84 \times 10^{-21}$ |
| Early Earth | 0.50 | $1.42 \times 10^{-3}$ | 0.50 | $1.18 \times 10^{-16}$ |
| Super-Earth | 0.50 | $1.06 \times 10^{-4}$ | 0.50 | 0.19 |
| Gliese 581c | 0.34 | $2.84 \times 10^{-3}$ | 0.66 | 1099 |
| CoRoT-7b | 0.34 | $3.06 \times 10^{-6}$ | 0.66 | 1.48 |

609
610
611